\documentclass[pre,twocolumn,showpacs]{revtex4}
\usepackage{amsmath,amssymb}
\usepackage{epsfig}

\begin{document}

\newlength{\mylen}
\setlength{\mylen}{\textwidth}
\addtolength{\mylen}{-1cm}
\newcommand{\bea}{\begin{eqnarray}}
\newcommand{\eea}{\end{eqnarray}}
\newcommand{\vect}[1]{\mathbf{#1}}
\newcommand{\eps}{\epsilon}

\title{Multipole expansion of the electrostatic interaction  between charged colloids at interfaces}

\author{A. Dom{\'i}nguez$^1$, D. Frydel$^2$\ and M. Oettel$^3$}
\thanks{Supported by the German Science Foundation (DFG) via the Collaborative
Research Center SFB--TR6 ``Colloids in External Fields'', project section D6.}
%\email{oettel@mf.mpg.de}
\affiliation{$^1$F\'\i sica Te\'orica, Universidad de Sevilla, Apdo.1065, E-41080
  Sevilla, Spain}
\affiliation{$^2$Max--Planck--Institut f\"ur Metallforschung, Heisenbergstr.~3, D--70569 Stuttgart, Germany, and }
\affiliation{Institut f\"ur Theoretische und Angewandte
  Physik, Universit\"at Stuttgart, Pfaffenwaldring 57, D--70569 Stuttgart, Germany}
\affiliation{$^3$Johannes Gutenberg Universit\"at Mainz, Institut f{\"ur} Physik,
  WA 331, D--55099 Mainz, Germany}

\begin{abstract}
 The general form of the electrostatic potential around an arbitrarily charged 
colloid at a flat interface
between a dielectric and a screening phase (such as air and water, respectively) is
analyzed in terms of a multipole expansion. The leading term is isotropic in the
interfacial plane and varies with $d^{-3}$ where $d$ is the in--plane distance
from the colloid. The effective interaction  potential between two arbitrarily 
charged colloids is likewise isotropic
and $\propto d^{-3}$, thus generalizing the dipole--dipole repulsion first
found for point charges at water interfaces. Anisotropic, attractive interaction terms can arise
only for higher powers $d^{-n}$ with $n \ge 4$. 
The relevance of these findings for recent experiments is discussed.

\end{abstract}

\pacs{82.70.Dd}
 
\maketitle

\section{Introduction}

%{\em Introduction.--} 
The self--assembly of stably trapped,
sub-$\mu$m colloidal particles at water--air or water--oil interfaces
has gained much interest in recent years. 
%These particles are trapped irreversibly 
%at the interface if water wets the colloids only partially. 
For the
specific case of charge--stabilized colloids at interfaces, the {\em repulsive} part of their  mutual
interaction resembles a dipole--dipole interaction at large
separations.
This may be understood theoretically by approximating the colloid
as a point charge located either right at the interface \cite{Stil61,Hur85}
(i.e. assuming charges on the colloid--water interface)
and/or above the interface \cite{Ave00a} (i.e. assuming charges on the colloid--air/oil 
interface).
Additionally, the formation of metastable mesostructures with such colloids point 
to the possible existence of intercolloidal {\em attractions} far
beyond the range of van--der--Waals forces \cite{Gom05,Che05,Che06}, however,
care must be taken to avoid contaminations of the interface which lead
to colloid mesostructures with similar appearance \cite{Fer04}. 
%According to
%Refs.~\cite{Ghe97,Ghe01,Gar98a,Gar98b,Sta00,Que01,Gom05}, polystyrene spheres
%(radii $R=0.25\dots 2.5$ $\mu$m)
% on flat water--air interfaces  using deionized water exhibit
%spontaneous formation
%of complicated metastable mesostructures. They are consistent with the presence
%of an attractive, secondary
%minimum in the effective intercolloidal potential at separations $d/R\approx 3\dots 20$ with
%a depth of a few $k_B T$.
Previous work \cite{For04,Oet05,Oet05a,Wue05,Dom07} aimed at relating this attractive 
minimum to capillary interactions
due to interfacial deformations caused by a {\em homogeneous} surface charge on the
colloids but with no conclusive answer.
In recent work \cite{Che05,Che06},
it was experimentally shown that the charge--carrying surface groups used
for charge--stabilizing polystyrene colloids are actually distributed quite
inhomogeneously and patchily over the colloid surface. Thus it was
speculated in Refs.~\cite{Che05,Che06} that through this inhomogeneous
charge distribution like--charged colloids could acquire effective dipole
moments in the interface plane and  attractive electrostatic interactions of dipole--dipole
type could arise which might overcome the repulsion at shorter distances.

Motivated by the finding of inhomogeneous surface charge on colloids, we
extend the asymptotic results for the electrostatic potential and interaction of
point charges at water interfaces \cite{Hur85} to the general case of an
arbitrary, localized colloidal charge distribution using a multipole expansion. 
The presence of the interface leads to restrictions in the 
multipole coefficients of the potential around a single colloid and of the interaction energy
between two colloids. In particular, we
find that the leading term in the effective interaction energy between two colloids at
lateral distance $d$ is isotropic in the
interfacial plane, repulsive and $\propto
d^{-3}$ regardless of the inhomogeneities of the charge distribution in the colloids.
% if the colloids are spherical and inhomogeneously charged. 
Angular dependencies enter the effective interaction potential only in higher orders. %, starting from $d^{-4}$. 

\section{Electrostatics at water interfaces}

%\label{sec:water}

\subsection{A toy model: water as a perfect conductor}

\begin{figure}
 \begin{center}
  \epsfig{file=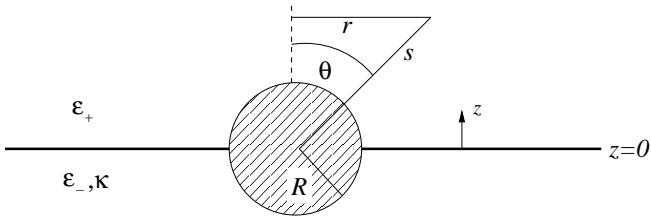, width=\columnwidth}
 \end{center}
 \caption{Geometrical configuration of the electrostatic problem: the
   potential %$\Phi(\vect r)$
   is calculated in the domain outside the sphere of radius $R$, which
   encloses the colloidal particle. The flat interface at $z=0$
   separates an upper dielectric phase (dielectric constant $\eps_+$)
   from a lower electrolytic phase (dielectric constant $\eps_-$,
   Debye screening length $\kappa^{-1}$).}
 \label{fig:geom}
\end{figure}

%{\em A toy model: water as a perfect conductor.--} 
For a quick insight on the
effect of an interface on the multipole expansion of the electrostatic potential,
we consider the water phase being a perfect conductor.
%, i.e.the Debye screening length $\kappa^{-1} \to 0$. 
The flat interface is located at
$z=0$ and the colloid is modelled by
a fixed charge distribution $\rho_{\rm C}(\vect r)$ above the water phase. %which is zero inside the water phase ($z<0$).
The boundary condition at $z=0$ simply implies that there 
is no tangential (or in--plane) electric field and the potential for $z>0$ can be obtained with the method of image charges.
% results from a charge configuration made of the colloid charge and a mirror image charge
% of opposite sign in the water phase. 
Therefore, the effective (real + image) charge distribution is
spatially localized and can be enclosed in a ball of finite radius $R$
(see Fig.~\ref{fig:geom} with $\kappa^{-1} \to 0$). In standard
spherical coordinates $(s,\theta,\varphi)$ measured from the center of
this ball, the potential in the upper phase for $s>R$ can be written
as a multipole expansion (in the remainder of the paper, the $+$($-$) index
will refer to evaluation in the upper(lower) phase):
% We use standard spherical coordinates [$(x,y,z)=s\,(\sin\theta\cos\varphi,\,
% \sin\theta\sin\varphi,\, \cos\theta)$] and unnormalized spherical harmonics
% $Y_{lm}(\theta,\varphi)=P_l^m(\cos\theta)\,e^{im\varphi}$ with $P_l^m(x)$ denoting the associated
% Legendre polynomials.
% The multipole expansion for the potential in phase I ($z>0$) exists since
% the charge and image charge distribution is localized and it reads:
\bea
 \label{eq:mp}
  \Phi_+ (s,\theta,\varphi) %\sim \sum_{lm} \Phi_{{\rm I},lm} 
   = \sum_{\ell m} a_{\ell m} s^{-\ell-1} Y_{\ell m}(\theta,\varphi)\;,
\eea
in terms of normalized spherical harmonics $Y_{\ell m}$.
The boundary condition of vanishing in--plane electric field at the interface ($\theta=\pi/2$)
%($\propto  s^{-l-2}\, Y_{lm}(\pi/2,\varphi)$) 
implies 
$a_{\ell m}=0$ for $\ell+m$ even. Thus, the monopole vanishes ($a_{00}=0$) 
as well as the in--plane dipole ($a_{1\,\pm 1}=0$), and the leading decay is described generically by a dipole perpendicular to the interface ($a_{10}\neq 0$). 
% The off--diagonal quadrupole
% corresponding to $a_{2\,\pm 1}$ is in general non-zero but could be reduced to zero
% by a specific choice of the origin and the in-plane orientation of the coordinate
%system. The electrostatic interaction energy of the colloid with a second one
% located at an in--plane distance $\vect d=(d_x,d_y)=d(\cos\varphi,\sin\varphi)$ and described by a charge density  
% $\rho_{{\rm C},2}(\vect r)$ is given by
Consider a second, identical colloid located at an in--plane position
$\vect d=(d_x,d_y)$. %=d(\cos\varphi,\sin\varphi)$.
The total potential $\Phi_+$ is now the linear superposition of the
single--particle potentials $\Phi_+^{0}$ by each colloid, and the
electrostatic energy of the two--particle configuration is
\bea
% \label{eq:Ugen}
  U = U^{0} + \int d^3 r\, \rho_{\rm C}(\vect r) \, 
  \Phi_+^{0} ({\vect r}+{\vect d})\;,
\eea
where $U^{0}$ is the energy in the limit $d\to\infty$.
%(``self--energy'' of the colloids). 
Taylor expanding $\Phi_+^{0}$ about ${\vect r}={\vect 0}$ one obtains to
leading order in $1/d$
\bea
\label{eq:Utoy}
U - U^{0} \sim \frac{p_{z}^2}{2 d^{3}} \qquad (d\to\infty) ,
\eea
where $p_z =2 \int d^3 r \, z\, \rho_{\rm C}(\vect r) = a_{10}
\sqrt{3/(4\pi)}$ is the dipolar moment of $\rho_{\rm C}$ and its image
charge in the direction normal to the interface.
 This dipole--dipole interaction energy differs by a factor of one-half from the textbook result because 

\subsection{Water as a conductor with linear screening}

%{\em Water as a conductor with linear screening.--} 
The 
image charge construction in the case of perfectly conducting water
provides an intuitive explanation of 
 the origin of the normal dipole
and the absence of an in--plane dipole.
In the following we demonstrate that this finding still holds in the
more realistic case of water being an electrolyte and the
colloidal particle having an arbitrary shape, possibly protruding into
the region $z<0$, with given charge distribution and dielectric
properties.
Assuming linear screening, the electrostatic potential satisfies $(\Delta
-\kappa_\pm^2)\Phi_{\pm}(\vect r)=0$ with $\kappa_+=0$ and
$\kappa_-=\kappa$ being the inverse screening length in water. Using
standard cylindrical coordinates $(r,z,\varphi)$, we search for a solution outside
a ball of radius $R$ whose center is the coordinate origin and which encloses the colloid 
(see Fig.~\ref{fig:geom})
with the
boundary conditions that the potential (i) vanishes at infinity, (ii)
reduces to a given potential $\Phi_R(\theta,\varphi)$ at the surface of the ball
$s=R$, (iii) is continuous at the interface $z=0$, and (iv) that the associated
electric displacement perpendicular to the interface is continuous,
i.e.,
\bea
\label{eq:electricD}
\eps_+ \left.\frac{\partial\Phi_+}{\partial z}\right|_{z=0} & = &
\eps_- \left.\frac{\partial\Phi_-}{\partial z}\right|_{z=0}  
\qquad (r>R) .
\eea
The function $\Phi_R(\theta,\varphi)$ is determined by the
solution of the electrostatic problem inside the ball and
contains the relevant information on the precise geometrical and
electric properties of the particle.
By decomposing the problem in the full domain into the solution of
problems in simpler domains (the exterior of the sphere $s=R$ and each
of the halfspaces defined by $z=0$ (details can be found in
App.~\ref{sec:app}), one can finally write the solution as the
superposition $\Phi_\pm (\vect r) = \Phi^{\rm cyl}_\pm (r,z,\varphi) +
\Phi^{\rm sph}_\pm (s,\theta,\varphi)$, where the contribution
$\Phi^{\rm cyl}_\pm (r,z,\varphi)$ (using cylindrical coordinates) is given by
\bea
 \label{eq:cyl}
 \Phi^{\rm cyl}_\pm (r,z,\varphi) &=& \sum_{m=-\infty}^{+\infty} e^{i m\varphi}
 \int_0^\infty \!\!\! dq\, A_m (q) J_{|m|}(q r) \, {\rm e}^{- K_{\pm}z} \nonumber \\
%  & & \left( K_\pm=\pm \sqrt{q^2+\kappa_\pm^2} \right) ,
  %(K_+=-q,  \quad K_{-}=\sqrt{q^2+\kappa^2})
\eea
with $K_\pm=\pm \sqrt{q^2+\kappa_\pm^2}$,
and the contribution $\Phi^{\rm sph}_\pm$ (using spherical coordinates) reads
\bea
 \label{eq:sph}
 \Phi^{\rm sph}_\pm (s,\theta,\varphi) &=& \sum_{\ell=0}^\infty \sum_{m=-\ell}^{+\ell} 
 C_{\ell m}^\pm {\cal R}_\ell^\pm (s) Y_{\ell m}(\theta,\varphi) \nonumber \\
 & & \left( {\cal R}_\ell^\pm (s) = s^{\ell} \frac{d^\ell}{(s\, ds)^\ell} \left[\frac{{\rm e}^{-\kappa_\pm s}}{s} \right] \right).
\eea
The coefficients $C_{\ell m}^\pm$ are given by
\bea
\label{eq:Ccoeff}
%  C_{\ell m}^\pm & = & \pm [1 - (-1)^{\ell-m}] 
%  \int_0^{2\pi} d\varphi \int_{(1\mp 1) \pi/4}^{(3\mp 1)\pi/4} d\theta\, \sin\theta 
%  \, Y_{\ell m}^* (\theta,\varphi) \times\mbox{} \nonumber \\
%  & & \mbox{}\times \left[ \Phi_R (\theta,\varphi) - \Phi_{\pm}^{\rm cyl} (r=R\sin\theta, z=R\cos \theta,\varphi) \right] ,
  C_{\ell m}^\pm & = & [1 - (-1)^{\ell-m}] 
  \int_0^{2\pi} d\varphi \int_{0}^{\pm 1} d(\cos\theta) 
  \, Y_{\ell m}^* (\theta,\varphi) \times\mbox{} \nonumber \\
  & & \mbox{} \left[ \Phi_R (\theta,\varphi) - \Phi_{\pm}^{\rm cyl} (r=R\sin\theta, z=R\cos \theta,\varphi) \right] , \qquad
\eea
such that $\Phi^{\rm sph}_\pm =0$ at $z=0$, $\Phi_\pm (s=R) = \Phi_R
(\theta,\varphi)$, and the boundary conditions (i)--(iii) are satisfied
automatically.
The coefficients $A_m(q)$ in the expression for
$\Phi_\pm^{\rm cyl}$ (Eq.~(\ref{eq:cyl})) must be chosen to enforce the
boundary condition~(\ref{eq:electricD}). This condition can be
extended to the region $0<r<R$ by continuing the fields
$\Phi_\pm(\vect r)$ with any virtual solution into the interior of the ball, $s<R$.
The precise form of the continuation is irrelevant, since the solution
outside the ball depends only on the potential at the surface of the ball, 
$\Phi_R(\theta,\varphi)$  (Faraday's cage effect).
% This also means that the solution for $A_m(q)$ is not unique, but
% depends on the arbitrary continuation.
Thus, by using orthonormality and closure of the set of
Bessel functions $\{J_{|m|}(q r)\}_{q\in [0,\infty)}$ in the domain
$0<r<\infty$, Eq.~(\ref{eq:electricD}) can be solved for the
coefficients $A_m(q)$:
\bea
\label{eq:Acoeff}
A_m(q) & = & \frac{q \,\sum_{\ell=0}^\infty \left[ 
    \eps_+ \hat{C}^+_{\ell m} \gamma^+_{\ell m}(q) - 
    \eps_- \hat{C}^-_{\ell m} \gamma^-_{\ell m}(q) 
  \right]}{\eps_+ q + \eps_-\sqrt{q^2+\kappa^2}} , \qquad
\eea
with %the coefficients
$
\hat{C}^\pm_{\ell m}  :=   - {\rm e}^{-i m\varphi}
\partial_\theta Y_{\ell m}(\theta=\pi/2,\varphi) \, C^\pm_{\ell m} 
$
and %the functions
$
  \gamma^\pm_{\ell m} (q) := \int_R^{+\infty} dr\; {\cal R}^\pm_\ell(r)\, J_{|m|}(qr) 
 $, 
which are the Hankel transforms of the radial dependence of the
spherical part $\Phi_\pm^{\rm sph}$ (see Eq.~(\ref{eq:sph})) continued into the region
$s<R$ by zero. Eq.~(\ref{eq:Acoeff}) is not the explicit expression
for the coefficients $A_m(q)$ because they appear implicitly also in
the coefficients $C^\pm_{\ell m}$, see Eq.~(\ref{eq:Ccoeff}), but it
does provide their dependence on $q$. In particular, for $\ell-m$ odd
(i.e., when $C^\pm_{\ell m}\neq 0$), the functions
$\gamma^\pm_{\ell m}(q)$ possess a Taylor expansion around $q=0$ with the
lowest term being of order $q^{|m|}$, so that %as $q\to 0$ one has
\begin{equation}
  \label{eq:Aexp}
  A_m(q) = \sum_{j=0}^\infty a_{jm} q^{j} , \qquad  
  \textrm{with $a_{jm}=0$ if } j\leq |m| .
\end{equation}
The existence of a Taylor expansion in $q$ of the coefficients $A_m(q)$
allows to extract the large--$r$ behavior of the potential and the $z$--component of the
 electric field at the
interface. Introducing the factors
\bea
  {\cal J}_{jm} & := & \lim_{z\to 0} \int_0^\infty dp \, p^j J_{|m|}(p) {\rm e}^{-z p} 
  = \frac{2^j\, \Gamma\left(\frac{|m|+j+1}{2}\right)}{\Gamma\left(\frac{|m|-j+1}{2}\right)} \nonumber
%   = \left\{
%     \begin{array}[c]{cl}
%       0 & \textrm{if $j-|m|$ odd and $j-|m|\geq 1$}, \\
%       & \\
%       \neq 0 & \textrm{if $j-|m|$ even or $|m|>j$} .
%     \end{array}\right.
\eea
and inserting the expansion~(\ref{eq:Aexp}) into the
corresponding definitions of the fields, one 
obtains \footnote{These are asymptotic
  expansions. There are also exponentially decaying terms which cannot
  be recovered from an expansion like Eq.~(\ref{eq:Aexp}).}
\bea
  \Phi_\pm(r,\varphi,z=0) & \sim & \sum_{j=0}^\infty\;\sum_{m=-j+1}^{j-1} 
   \frac{{\rm e}^{i m \varphi}}{r^{j+1}} \, a_{jm} {\cal J}_{jm} ,  \\
  & \sim & \mbox{} - \frac{a_{20}}{r^3} - 
  \frac{3}{r^4} \sum_\pm a_{3\pm 1} {\rm e}^{\pm i \varphi} \cdots ,
  %\left[ a_{31}{\rm e}^{i \varphi} + a_{3-1}{\rm e}^{-i \varphi}\right] \cdots , %O(1/r^5) ,
  \nonumber 
\eea
\bea
 \left.\frac{\partial\Phi_+}{\partial z}\right|_{z=0} 
 & \sim & \sum_{j,m}
% \sum_{j=0}^\infty \sum_{m=-j+1}^{j-1} 
  \frac{{\rm e}^{i m \varphi}}{r^{j+2}} \, \left[ r^{j+1} {\cal R}^+_j(r) \hat{C}^+_{jm} 
 - a_{jm} {\cal J}_{j+1,m}\right] , \nonumber \\
 & \sim &
 \frac{a_{10} + \hat{C}^+_{10}}{r^3} + 
 \frac{3}{r^4} \sum_\pm (a_{2\pm 1}+\hat{C}^+_{2\pm 1}) {\rm e}^{\pm i \varphi} \cdots , \nonumber 
 %\left[ (a_{21} + \hat{C}^+_{21}){\rm e}^{i \varphi} \right. \nonumber \\
 %& & \left. \mbox{} + (a_{2-1} + \hat{C}^+_{2-1}) {\rm e}^{-i \varphi} \right] + \cdots , \nonumber
\eea
where we have used that $C^+_{jm}=0$ if $|m|=j$ and $\hat{C}^+_{jm}=0$ 
whenever ${\cal J}_{j+1,m}=0$.
Therefore, both the potential and the normal component of the electric field at the interface
are asymptotically dominated by an {\em angular--independent}
decay  $\propto 1/r^3$; anisotropic behavior arises only in subleading terms.
By continuity, this conclusion also holds asymptotically for the fields at a fixed
height $h$ above or below the interface ($r\gg|h|$).

This result is not exclusive of the single--particle configuration: if
there are several particles at the interface, one can surround each of
them by a ball of radius $R$ and the solution $\Phi(\vect r)$ of the
electrostatic problem will be written as a superposition of
single--particle potentials determined by the total potential at the
surface of each ball (in general different from the potential
$\Phi_R(\theta,\varphi)$ in the single--particle configuration).  For
each of these single--particle potentials the expansion~(\ref{eq:Aexp})
still holds, since it does not depend on the precise value of
the potential at the balls. \\

\subsection{An illustrative 2d example}

\begin{figure}
 \begin{center}
  \epsfig{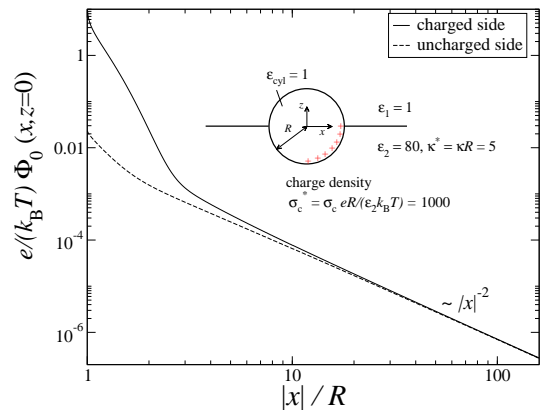}
 \end{center}
 \caption{Potential along the interface for an inhomogeneously charged cylinder half--immersed in water.
Maxwell's equations hold in the upper phase and the cylinder, the Poisson--Boltzmann equation
holds in the lower phase.
The parameters are given in the inset, the numerical calculations have been done using
FEMLAB.}  
 \label{fig1}
\end{figure}

%{\em An illustrative 2d example:} 
We calculated the electrostatic
potential for an inhomogeneously charged cylinder at an air--water interface 
(see the inset of Fig.~\ref{fig1} for some definitions). Because of its two--dimensional nature,
this problem is amenable to a numerical treatment. Here, the multipole expansion at the interface
gives $\Phi_\pm (z=0) \sim a_0 \ln |x| + p_x/x + q_{xx}/x^2 + \dots $ and $\left.\partial \Phi_+/\partial z\right|_{z=0} \sim  p_z/x^2 + \dots$.    
The numerical solution for $\Phi_\pm$ shows
that the in--plane dipole term $\propto x^{-1}$ is absent and the asymptotic expansion 
starts with the quadrupolar term (see Fig.~\ref{fig1}).
The asymptotics for $\left.\partial \Phi_+/\partial z\right|_{z=0}$ (not shown) also
contains the term $\propto x^{-2}$, which is interpreted as the effect of a counter--ion generated
dipole component $p_z$ perpendicular to the interface. These findings, most notably
the absence of $p_x$, match the previous ones in three dimensions. \\

\subsection{The effective interaction energy}

The free energy functional of a multiparticle configuration in the
linear screening approximation reads \cite{ShHo90} 
\bea
F[\Phi] = \int d^3 r\, \left[ \rho_{\rm C}({\vect r}) \Phi %({\vect r})
  - \frac{ \eps({\vect r})}{8\pi} \left[ 
    \kappa^2({\vect r}) \Phi^2 + |\nabla\Phi|^2 \right]
\right] ,
\eea
where the charge density $\rho_{\rm C}({\vect r})$ is localized on the
colloidal particles. This includes the electrostatic energy as well as the entropy associated to the ion distribution.
The extremum of $F[\Phi]$ provides the field equation in thermal
equilibrium, $\nabla\cdot[\eps\nabla\Phi] = \eps\kappa^2\Phi -
4\pi\rho_C$. With the help of this equation, the free energy in
equilibrium simplifies to
\bea
\label{eq:totalF}
F_{\rm eq}(\{ {\vect x}_\alpha \}) = 
\frac{1}{2} \int d^3 r\, \rho_{\rm C}({\vect r}) \Phi ({\vect r}) ,
\eea
which is known as the ''potential of mean force'' for the degrees of
freedom ${\vect x}_\alpha$ (position of the center of a ball of
radius $R$ enclosing the $\alpha$-th particle).
One may decompose 
$F_{\rm eq} = F^{0} + \delta F$, where $F^{0}$ is the equilibrium free
energy in the limit $|{\vect x}_\alpha - {\vect x}_\beta|\to\infty$
(isolated particles).
The total potential can be similarly written as $\Phi = \sum_\alpha
\Phi^{0}_\alpha + \delta\Phi$, where $\Phi^{0}_\alpha({\vect r})$
denotes the potential field generated by the $\alpha$-th particle in
isolation and $\delta \Phi({\vect r}; \{{\vect x}_\alpha\})$ is the
total perturbation induced by the presence of other particles. Due to
the linear nature of the problem, the perturbation $\delta \Phi$,
\bea
  \delta \Phi ({\vect r}) = \sum_{\alpha\neq\beta} \int_{|{\vect r}'-{\vect x}_\alpha|<R} d^3 r' \, 
  G_{\alpha\beta} ({\vect r}, {\vect r'}) 
  \Phi^{0}_\beta ({\vect r}') ,
\eea
can be written
in terms of a generalized %, operator--valued
susceptibility $G_{\alpha\beta}({\vect r}, {\vect r}')$ depending on
the precise shape and charge distribution of the particles.

Since $\Phi^{0}({\vect r})$ near the interface exhibits asymptotically
an isotropic decay $\propto 1/r^3$,
$\delta \Phi({\vect r}; \{{\vect x}_\alpha\})$
depends {\em only} on $d_{\alpha\beta}= |{\vect x}_\alpha - {\vect
  x}_\beta|$ (and not on the orientation of
${\vect x}_\alpha - {\vect x}_\beta$) in the asymptotic limit
$d_{\alpha\beta}\to\infty$.
Furthermore, $\delta \Phi$ is rescaled by a factor
$\lambda^{-3}$ if all distances $d_{\alpha\beta}$ are rescaled
simultaneously by a factor $\lambda$.
From Eq.~(\ref{eq:totalF}) 
the same property holds for $\delta F (\{{\vect x}_\alpha\})$.
In particular, for a
two--particle configuration this yields an asymptotic 
potential of mean force of the form
\bea
\label{eq:U2part}
 F_{\rm eq} (d) - F^0 \sim \frac{B}{d^3} \qquad (d\to\infty) ,
\eea
and the constant $B$ is positive for like particles.
In analogy with Eq.~(\ref{eq:Utoy}), it is natural to
interpret this expression as the interaction energy 
between two effective dipoles perpendicular to the interface.

\section{Discussion and Conclusion}

%{\em Discussion and Conclusion.--}
We have shown that the form of the multipole expansion of the potential around a charged colloid
and of the effective interaction energy between two colloids trapped at a water interface
is qualitatively different
%leads to certain restrictions in powers and orientation dependence of the multipole terms
from the situation in bulk. 
The dominating interaction terms can be qualitatively understood by assuming 
water to be a perfect conductor.
%If we assume the colloidal shape to be symmetric with respect to
%rotations around the interface normal,
The leading interaction term between the colloids a distance $d$ apart is of dipole--dipole type
($\propto d^{-3}$) and isotropic in the 
interfacial plane.
In other words, even if the charges on the colloid surface are distributed
arbitrarily
the counterions arrange themselves such that asymptotically
the configuration corresponds to an effective dipole perpendicular to the interface.
Orientation--dependent interactions and thus possible attractions for like--charged
colloids only arise
in subleading order.
 
This is in marked contrast to the analysis of the experiment reported in 
Refs.~\cite{Che05, Che06}.
Motivated by the experimentally found inhomogenous surface charge, it was
pictorially suggested (see Fig.~1 in Ref.~\cite{Che05}) 
that spontaneous fluctuations in the colloid's orientation would generate
(via an instantaneously equilibrating counterion cloud)
effective in--plane dipoles $\vect p_i$ with corresponding interactions
$\propto [d^2 ({\vect p}_1\cdot{\vect p}_2) -3 ({\vect d}\cdot {\vect p}_1)({\vect d}\cdot {\vect p}_2)]/d^5$.
%\bea
% \label{eq:ufluc}
% U_{\rm fluc}(d) \propto  \frac{  -
%  3 ({\vect d}\cdot {\vect p}_1)({\vect d}\cdot {\vect p}_2)}{d^5}
%\eea
After averaging over the orientation fluctuations, such an interaction would lead to
an effective, isotropic attraction competing with the isotropic dipole--dipole
repulsion. According to the model worked out by the authors, the total interaction potential would exhibit an attractive minimum due to
the effect of the fluctuating in--plane dipoles at rather small distances ($d \simeq 2.2$ colloid radii $R_{\rm C}$, so small that already the use of a pure dipole--dipole interaction casts serious doubts on the reliability of the model).
The analysis in Ref.~\cite{Zho07} purported to support this picture is
actually incomplete and just states that no monopolar term arises,
without entering into a systematic analysis of constraints on higher
order multipoles. In another note \cite{Zho07a} the existence of
the Taylor expansion of the coefficients $A_m(q)$ around $q=0$ (see
Eq.~(\ref{eq:Aexp})) was doubted on which the asymptotic analysis of the 
electrostatic potential and field is based. The present explicit 
proof of the analyticity of the coefficients $A_m(q)$ should disperse such doubts.

The results of our work imply that
asymptotically an in--plane dipolar interaction  cannot arise if the
counterions are equilibrated (see Eq.~(\ref{eq:U2part})). Consequently one cannot
expect asymptotically relevant attractions from the orientational fluctuations of the
colloids. However, for smaller $d$ the asymptotic $1/d$ expansion is likely to break
down. For small colloid radius, $R_{\rm C} \ll \kappa^{-1}$, this becomes relevant when
$d \sim \kappa^{-1}$: in this case the screening clouds of the colloids overlap
and the interaction falls off exponentially with $d$ before crossing over to the algebraic decay \cite{Hur85,Dom07}. For large
colloid radius, $R_{\rm C} \gg \kappa^{-1}$, the precise shape and charge distribution of the colloids will determine the interaction whenever $d\sim R_{\rm C}$.
% effect of the ``holes" dug out by the
% colloids become relevant if the shortes distance between the colloid surfaces $d-2R \sim R$
% (see the discussion at the end of Sec.~\ref{sec:water}).  
Certainly, for both regimes a more elaborate numerical analysis of the anisotropy
in the electrostatic interactions is required to assess whether fluctuations in the 
orientation of the colloids may lead to attractions. 
However, even in that case their existence is doubtful looking at the general
results on the absence of like--charge attraction in confined geometries \cite{Tri99}. 
%  In any case, the thermal
% average will again lead to isotropy in the attractive terms, as opposed to the
% claims of orientation--sensitivity made in Refs.~\cite{Che05,Che06}.
In any case, the results from the model studied in
Refs.~\cite{Che05,Che06} are not reliable since the model presupposes
an interaction energy which does not satisfy the correct
asymptotic decay given by Eq.~(\ref{eq:U2part}).

%\vfill

\begin{appendix}

\section{Solution of the electrostatic problem}

\label{sec:app}

We consider the potential $\Phi_\pm({\vect r})$ in the domain shown in
Fig.~\ref{fig:geom} given as the solution to the following problem :
\begin{eqnarray*}
  \nabla^2 \Phi_+ & = & 0 , \qquad {\vect r} \in \{ s>R, z>0 \} , \\
  \nabla^2 \Phi_- & = & \kappa^2 \Phi_- , \qquad {\vect r} \in \{ s>R, z<0 \} , \\
  \Phi_+ (s=R, \theta, \varphi) & = & \Phi_R (\theta,\varphi) ,
  \qquad 0<\theta<\frac{\pi}{2} , \\
  \Phi_- (s=R, \theta, \varphi) & = & \Phi_R (\theta,\varphi) ,
  \qquad \frac{\pi}{2}<\theta<\pi , \\
  \Phi_+ (r,\varphi,z=0) & = & \Phi_- (r,\varphi,z=0), \qquad r>R , \\
  \epsilon_+ \left.\frac{\partial\Phi_+}{\partial z}\right|_{z=0} & = & 
  \epsilon_- \left.\frac{\partial\Phi_-}{\partial z}\right|_{z=0}, \qquad
  r>R , \\
  |\Phi({\vect r})| & < & \infty, \qquad |{\vect r}|\to\infty .
\end{eqnarray*}
Here, $\Phi_R(\theta,\varphi)$ is the potential at the surface of the
ball $s=R$ and is assumed to be given. In order to solve this problem,
we split it in two auxiliary problems,
one for each halfspace: \\

\begin{figure}[h]
  \begin{center}
    \epsfig{file=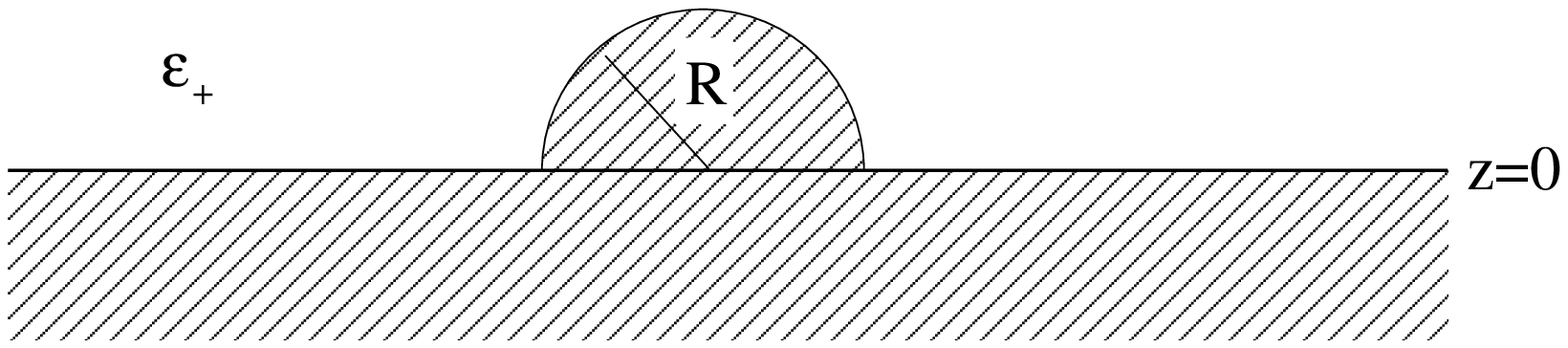, width=7cm}
  \end{center}
  \caption{Domain of definition of the problem {\bf UP}.}  
  \label{fig:geomUP}
\end{figure}
\noindent %
{\bf Problem UP} in the domain $s>R$ and $z>0$, see
Fig.~\ref{fig:geomUP}:
\begin{eqnarray*}
  \nabla^2 \Phi_+ & = & 0 , \qquad {\vect r} \in \{ s>R, z>0 \} , \\
  \Phi_+ (s=R, \theta, \varphi) & = & \Phi_R (\theta,\varphi) , 
  \qquad 0<\theta<\frac{\pi}{2} , \\
  \Phi_+ (r,\varphi,z=0) & = & F_0 (r,\varphi), 
  \qquad r>R , \\
  |\Phi_+ ({\vect r})| & < & \infty, \qquad |{\vect r}|\to\infty .
\end{eqnarray*}
\newline

\begin{figure}[h]
  \begin{center}
    \epsfig{file=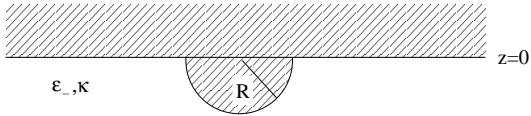, width=7cm}
  \end{center}
  \caption{Domain of definition of the problem {\bf LOW}.}  
  \label{fig:geomLOW}
\end{figure}
\noindent %
{\bf Problem LOW} in the domain $s>R$ and $z<0$, see
Fig.~\ref{fig:geomLOW}:
\begin{eqnarray*}
  \nabla^2 \Phi_- & = & \kappa^2 \Phi_- , \qquad {\vect r} \in \{ s>R, z<0 \} , \\
  \Phi_- (s=R, \theta, \varphi) & = & \Phi_R (\theta,\varphi) , 
  \qquad \frac{\pi}{2}<\theta<\pi , \\
  \Phi_- (r,\varphi,z=0) & = & F_0 (r,\varphi), 
  \qquad r>R , \\
  |\Phi_- ({\vect r})| & < & \infty, \qquad |{\vect r}|\to\infty .
\end{eqnarray*}
Here $F_0(r,\varphi)$ (=potential at the interface $z=0$) is an
auxiliary function which will be eventually determined by the boundary
condition~(\ref{eq:electricD}). Each of these problems can in turn be
decomposed in simpler problems, one with boundary conditions imposed
only at the plane $z=0$ and one with boundary conditions imposed only 
at the ball $s=R$: \\

\begin{figure}[h]
  \begin{center}
    \epsfig{file=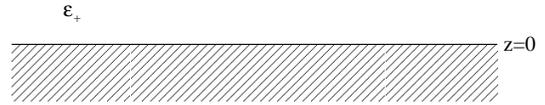, width=7cm}
  \end{center}
  \caption{Domain of definition of the problem {\bf UP-cyl}.}  
  \label{fig:geomUPcyl}
\end{figure}
\noindent %
{\bf Problem UP-cyl} in the domain $z>0$, see Fig.~\ref{fig:geomUPcyl}:
\begin{eqnarray*}
  \nabla^2 \Phi_+^{\rm cyl} & = & 0 , \qquad {\vect r} \in \{ z>0 \} , \\
  \Phi_+^{\rm cyl} (r,\varphi,z=0) & = & F_0 (r,\varphi) 
  + {\cal F}_0 (r,\varphi) , 
  \qquad 0< r , \\ 
  |\Phi_+^{\rm cyl} ({\vect r})| & < & \infty, \qquad |{\vect r}|\to\infty . 
\end{eqnarray*}
\newline
\begin{figure}[h]
  \begin{center}
    \epsfig{file=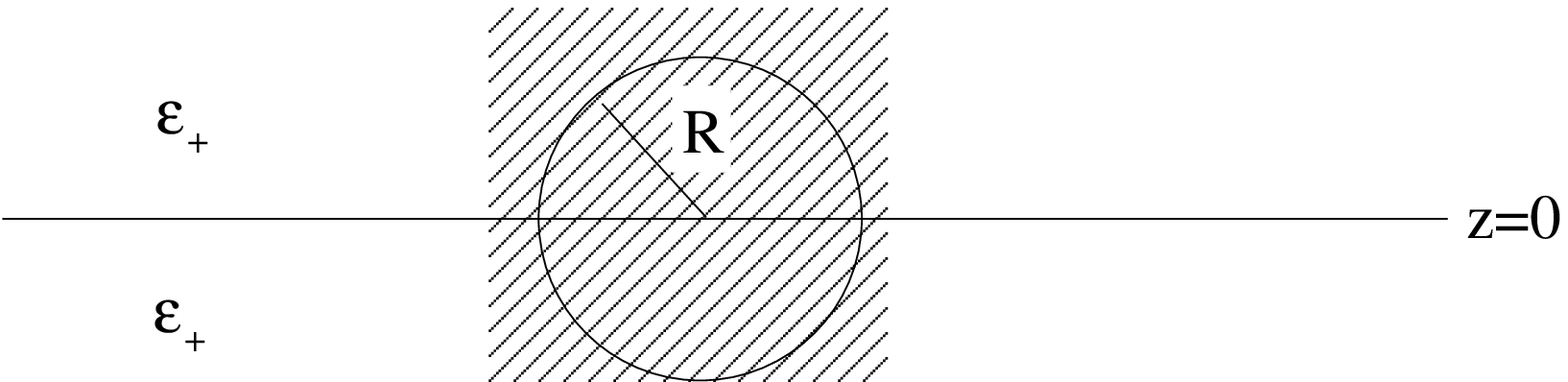, width=7cm}
  \end{center}
  \caption{Domain of definition of the problem {\bf UP-sph}.}  
  \label{fig:geomUPsph}
\end{figure}
\noindent %
{\bf Problem UP-sph} in the domain $s>R$, see Fig.~\ref{fig:geomUPsph}:
\begin{eqnarray*}
  \nabla^2 \Phi_+^{\rm sph} & = & 0 , \qquad {\vect r} \in \{ s>R \} , \\
  \Phi_+^{\rm sph} (R,\theta,\varphi) & = & \Phi_R (\theta,\varphi) - 
  \Phi_+^{\rm cyl} (R,\theta,\varphi) , 
  \quad 0 < \theta < \frac{\pi}{2} , \\
  \Phi_+^{\rm sph} (R,\theta,\varphi) & = & \mbox{} - 
  \Phi_+^{\rm sph} (R,\pi-\theta,\varphi) , 
  \qquad \frac{\pi}{2} < \theta < \pi , \\
  |\Phi_+^{\rm sph} ({\vect r})| & < & \infty, 
  \qquad |{\vect r}|\to\infty .
\end{eqnarray*}
Here the function ${\cal F}_0 (r,\varphi)$ verifies ${\cal F}_0
(r>R,\varphi)=0$ and is an otherwise arbitrary smooth function which
continues the potential $F_0 (r,\varphi)$ into the region $r<R$ of the
plane $z=0$. As discussed in the main text, ${\cal F}_0 (r,\varphi)$
is just an intermediary auxiliary function whose precise choice is
ultimately irrelevant for the determination of the total potential
outside the ball $s=R$.
With the choice of boundary condition for $\Phi_+^{\rm sph}$ at $s=R$
it is clear that $\Phi_+^{\rm sph}=0$ at $z=0$ and therefore
\begin{displaymath}
  \Phi_+ ({\vect r}) = \Phi_+^{\rm cyl} + \Phi_+^{\rm sph} , \qquad
  {\rm if}\quad {\vect r} \in \{ s>R, z>0 \}. 
\end{displaymath}
Analogously, the problem {\bf LOW} can be decomposed into a problem
{\bf LOW-cyl} and a problem {\bf LOW-sph} and
\begin{displaymath}
  \Phi_- ({\vect r}) = \Phi_-^{\rm cyl} + \Phi_-^{\rm sph} , \qquad
  {\rm if}\quad {\vect r} \in \{ s>R, z<0 \}. 
\end{displaymath}
Each of these simpler problems is now amenable to an analytical
solution, provided by Eqs.~(\ref{eq:cyl})--(\ref{eq:Ccoeff}). The
auxiliary functions $F_0$ and ${\cal F}_0$ are absorbed in the unknown
coefficients $A_m(q)$ in Eq.~(\ref{eq:cyl}), which are then determined
by Eq.~(\ref{eq:electricD}), the only boundary condition of the
original problem not taken into account by the stepwise process of
decomposing the problem into simpler ones.

\end{appendix}

\end{document}